\documentclass[seceq]{ptptex}

\usepackage{graphicx}

\preprintnumber[5cm]{ arXiv:1234.1234 [hep-ex] }
\title{ Search for the Higgs Boson and \\ Beyond Standard Model Phenomena in CMS }
\author{ Gr\'egory \textsc{SCHOTT}, \\ on behalf of the CMS collaboration }
\inst{ Institute of Experimental Nuclear Physics, Karlsruhe Institute of Technology }

\abst{ Results of recent Higgs boson and beyond standard model searches in CMS performed with datasets of $1.0-1.7\invfb$ will be summarized in this proceeding contributed to the $41^\mathrm{st}$ International Symposium on Multiparticle Dynamics (ISMD2011). }

\PTPindex{600}  % subject index(es): http://solution.dynacom.jp/cgi-bin/ptp/submission/subject_index.cgi

\newcommand\invfb{\ensuremath{\,\mathrm{fb}^{-1}}}
\newcommand\gev{\ensuremath{\,\mathrm{GeV}}}
\newcommand\tev{\ensuremath{\,\mathrm{TeV}}}
\newcommand\gevcc{\ensuremath{\,\mathrm{GeV}/c^2}}
\newcommand\too{\ensuremath{\rightarrow}}

\begin{document}

\maketitle

\section{Introduction}

The standard model (SM) of high-energy particle physics has been very efficient in describing the experimental measurements up to now but one of its key prediction has not yet been observed: the Higgs boson, a scalar neutral particle, which is at the source of the electro-weak symmetry breaking and provides a mechanism for particles to acquire mass. We know, however, that the SM theory breaks at larger scales and some other issues are also left open such as the unification of couplings, hierarchy problem, dark matter issue and neutrino masses. Theories have been proposed that attempt to fix some of those issues such as supersymmetry (SUSY) or other beyond standard models (BSM) and are currently being under experimental research. The CMS experiment, a multi-purpose detector \cite{cms_detector} operating at the CERN LHC $pp$ collider, has been designed to investigate for a wide range of physical phenomena. We will discuss in section \ref{sec:higgs} and section \ref{sec:bsm}, respectively, a selection of the latest Higgs and BSM searches, performed by the CMS collaboration with datasets of $1.0-1.7\invfb$. 

\section{Higgs Boson searches}
\label{sec:higgs}

The CMS collaboration searched for the Higgs boson in 8 channels with signatures involving isolated leptons or photons, jets, missing transverse energy (MET), $b$-tagging and $\tau$ identification. Because its production cross-section~\cite{lhc_xsec} is orders of magnitude smaller than that of dominating background processes, it is crucial in the analyses to reduce strongly and have a tight control on the remaining backgrounds.

\subsection{Low Mass Higgs}

At low mass, the Higgs has been searched in the channels $H\too\gamma\gamma$, $H\too\tau\tau$ and $H\too b\bar{b}$. The narrow peak signature obtained in $H\too\gamma\gamma$, thanks to excellent performance of CMS electromagnetic calorimeter, makes it a golden mode for the Higgs~\cite{hgg}. Large signal-like backgrounds, mostly from prompt QCD photons and fakes from jets, affect however the analysis. An optimization is performed, separating events in 8 categories based on conversion probability, $\eta$ and $p_T$ of the most energetic photon. Photon energy scale and resolution are extracted from $Z\too e^+e^-$ data events and backgrounds are determined from a fit to data.

In the $H\too\tau\tau$ analysis~\cite{htt}, channels with $\tau$ decaying into $\mu$, $e$ and hadrons: $\tau_e\tau_h$, $\tau_\mu\tau_h$, $\tau_e\tau_\mu$ have been studied and in particular the vector-boson fusion topology is exploited to improve the sensitivity to the Higgs.

\subsection{Intermediate and High Mass Higgs}

At intermediate and high mass, the channels $H\too WW\too 2l2\nu$, $H\too ZZ\too 4l$, $H\too ZZ\too 2l2\nu$, $H\too ZZ\too 2l2q$ and $H\too ZZ\too 2l2\tau$ have been studied. So far, the clearer SM Higgs exclusion is obtained in the $H\too WW\too 2l2\nu$ channel~\cite{hww}. Since, in addition to 2 isolated high-$p_T$ leptons, the signature involves also 2 missing $\nu$ there is large MET and no clear mass peak can be obtained. Kinematic observables are used to discriminate against backgrounds: jet multiplicity, $p_T$, $M_{ll}$, $M_T$ and $\Delta\phi(l^+l^-)$ (a small opening angle is expected in scalar Higgs boson decays).

$H\too ZZ\too 4l$ is the golden mode to search for the Higgs in the high mass region~\cite{hzz}. The events are fully reconstructed with mass resolutions of $2-4\gev$. Backgrounds are strongly reduced in the analysis. The irreducible $pp\too ZZ$ backgrounds is the almost exclusive one remaining and its rate, obtained from the $Z$ yield in data and theory for the ratio $\sigma_{ZZ}/\sigma_{Z}$, is predicted to be $21.2\pm0.8$. A shape analysis of the $M_{4l}$ spectrum is performed over the $21$ selected events.

\subsection{Statistical Combination of SM Higgs Searches}

\begin{figure}[h!]
  \centering
  \includegraphics[width=0.49\textwidth]{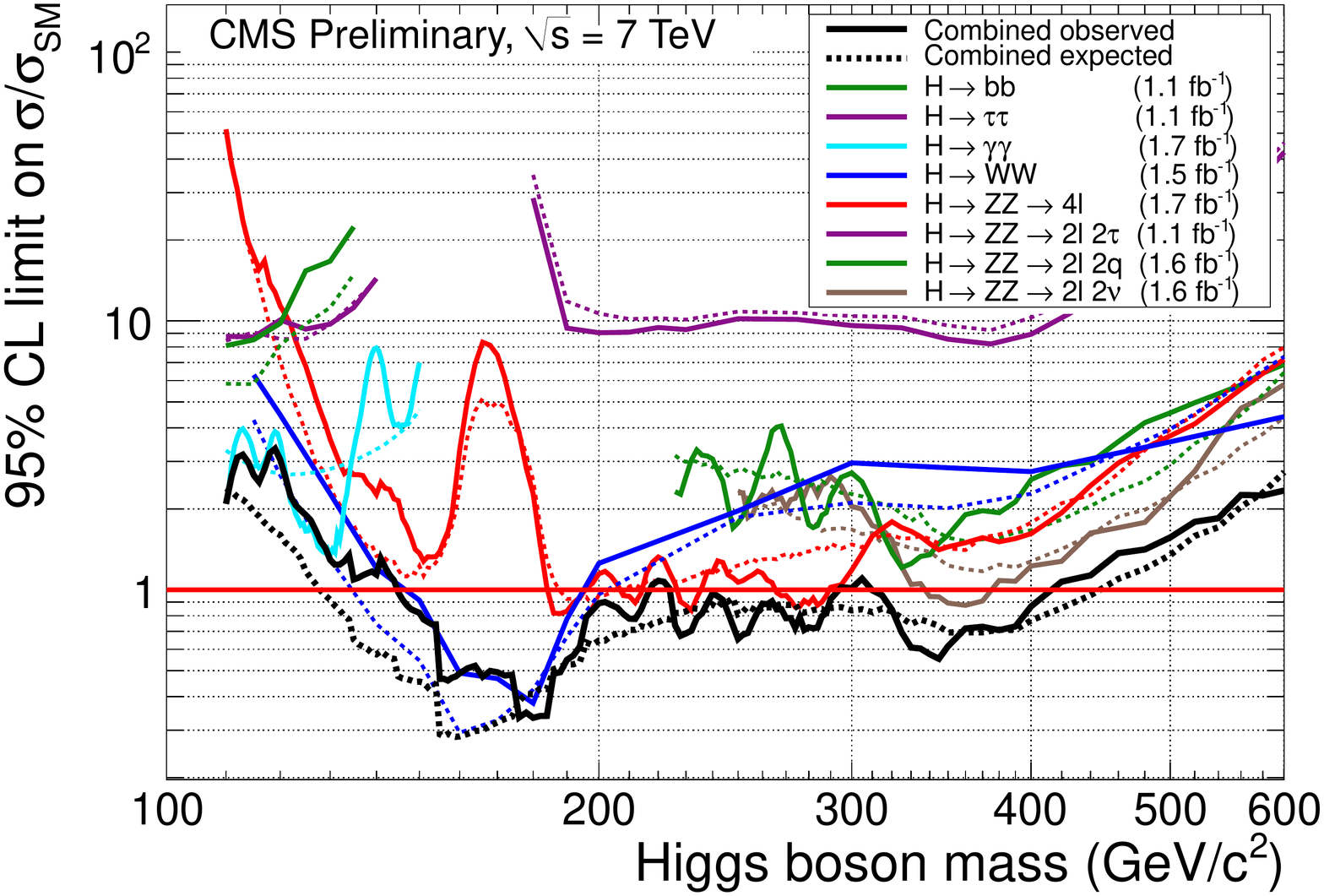}
  \includegraphics[width=0.49\textwidth]{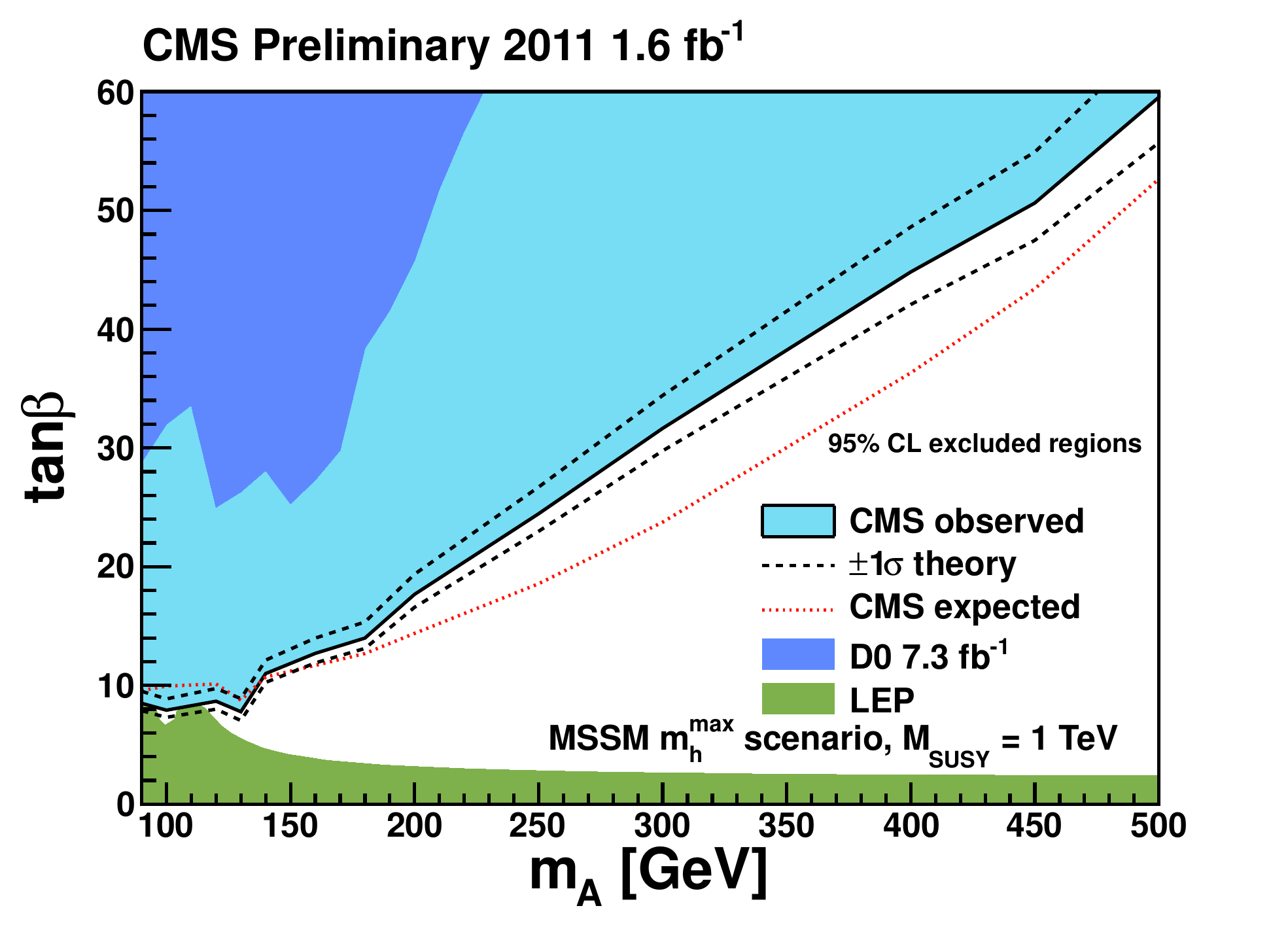}
  \caption{Left: 95\% CL exclusion limit on the Higgs production cross-section normalized by the SM-expected cross-section as function of the Higgs boson mass. Right: Constraints on the $m_A-\tan\beta$ plane from the MSSM Higgs analysis.
}
  \label{fig:higgs}
\end{figure}

Upper limits on $\mu=\sigma/\sigma_{SM}$ are computed in a $\mathrm{CL}_s$-based modified-frequentist approach that has been agreed across ATLAS and CMS experiments~\cite{lhc_hcg}. Nuisance parameters are taken into account through their profiling in the test statistic: $q_\mu=-2\ln[L(\mathrm{data};\mu,\hat\theta_\mu)/L(\mathrm{data};\hat\mu,\hat\theta_\mu)]$ with $0\leq\hat\mu\leq\mu$. Using the RooStats package, the mass ranges $145-216, 226-288, 310-400\gev$ are excluded at 95\% CL while the expected exclusion range for SM Higgs is $130-440\gev$ (see Fig.~\ref{fig:higgs})~\cite{hcombi}. Taking into account look-elsewhere effects, the probability of an excess at least as large as the one observed is evaluated to about $0.4$.

\subsection{Search for Non-Standard Model Higgs Bosons}

Constraints on the $m_A-\tan\beta$ plane from the MSSM $\Phi\too\tau\tau$ analysis are greatly improved (see Fig.~\ref{fig:higgs})~\cite{htt}. In analysis of $H^+\too \tau^+\nu$ in $t\bar t\too H^+W^-b\bar{b}$ with $t\too H^+ b$ upper limits of BR$(t\too H^+b)\lesssim4-5\%$ for $80<M_{H+}<160\gevcc$ are obtained~\cite{charged_higgs}. Also, exotic doubly-charged Higgs bosons produced in pairs are inclusively searched for in samples of 3 or more isolated leptons~\cite{doubly_charged_higgs}. While no signal excess is observed, the lower limits on $M_{\Phi^{++}}$ are improved from previous measurements and range from $269$ to $297\gev$, for chosen benchmark points of the type-II see-saw model.

\section{Beyond Standard Model}
\label{sec:bsm}

Many CMS analyses allow to search for manifestation of BSM phenomena including signatures for heavy resonances, strong gravity, long-lived particles or SUSY. Rather than investigate models one by one, the approach to study each signatures in a model-independent way allows also to cover for possible unexpected manifestations. 

One manifestation could be in the form of heavy resonances which are predicted by numerous models. This type of analysis require a good understanding of the detection performance of muons, electron, photons at high $p_T$ up to $1\tev$. New gauge bosons are searched in the channels $W'\too l\nu$ and $Z'\too l^+l^-$ ($l=e,\mu$). The $W'$ analysis is performed on the transverse mass distribution where the signal would manifest as an excess of very high mass events~\cite{w_prime} while in the $Z'$ analysis the dilepton invariant mass is used~\cite{z_prime}. When possible, data-driven control samples are used to constrain the background contributions. In both channels, good agreements are observed between expected and observed distributions. Assuming SM-like couplings, a counting analysis is performed to exclude $M_{W'}<2.3\tev$ at 95\% CL and, using the predicted mass shapes, $M_{Z'}<1.94\tev$. Limits of $M_{Z'_\psi}<1.62\tev$ and $M_{Z'(G_{KK})}<1.45-1.78\tev$ are obtained for a superstring-inspired $Z'_\psi$ and RSKK graviton (with couplings of $0.05-0.1$).
High-mass resonances are also searched with dijet events by looking for indications of a peak in a phenomenological fit of the exponentially-falling data distribution~\cite{dijet_resonance}. This analysis allows to probe for quark structure at high energies and yield 95\% CL lower limits on the mass for string resonances ($4.0\tev$), E$_6$ diquarks ($3.52\tev$), excited quarks ($2.49\tev$) or axigluons/colorons ($2.4\tev$) among others.

\begin{figure}[h!]
  \centering
  \includegraphics[width=0.60\textwidth]{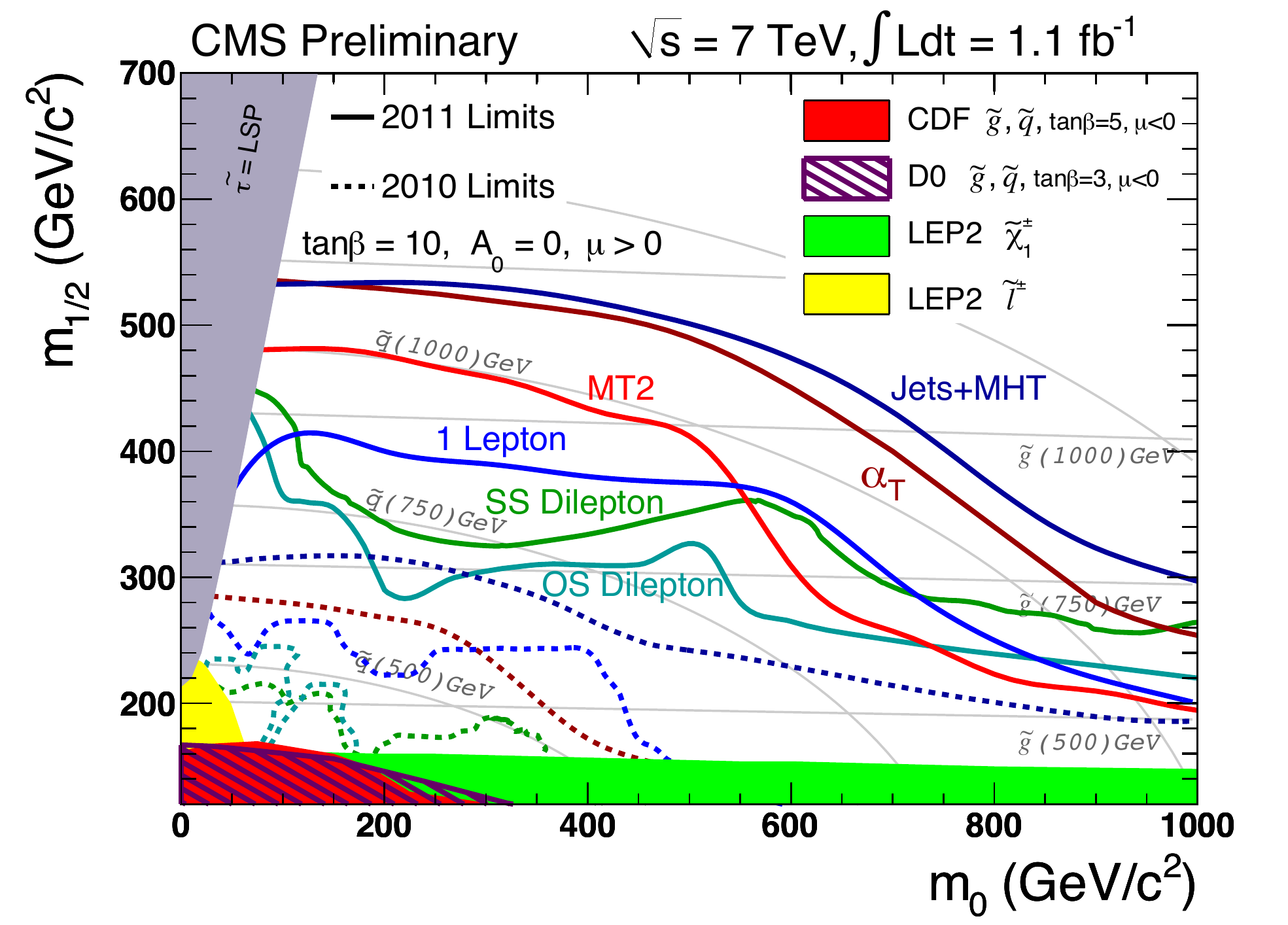}
  \caption{Summary of the observed limits in the CMSSM framework plotted in ($m_{0}$, $m_{1/2}$) plane.}
  \label{fig:susy}
\end{figure}

Multiple analyses sensitive to manifestations of SUSY and mainly dealing with final states of jets, leptons or photons and large missing transverse energy have been performed. Fig.~\ref{fig:susy} shows the constraints obtained in the CMSSM ($m_{0}$, $m_{1/2}$) plane greatly improving the former LEP2, Tevatron and LHC experimental results.

\section{Conclusion}

With the first inverse femtobarns of data acquired at the LHC experiments a very rich program opened up. A selection of recent CMS results has been summarized in this paper. In the SM Higgs sector, a combination of 8 analyses performed by the collaboration yields wide exclusion ranges disfavouring a medium mass Higgs while at lower masses, an excess is observed but will require further studies and increased sample sizes. Constraints on a $4^\mathrm{th}$ generation and fermiophobic as well as non-SM Higgs bosons have also been obtained. In the search for BSM phenomena, many types of manifestations have been sought but no evidence has yet been seen. Constaints on SUSY models have been improved and sensitivity to squarks and gluinos in range of $0.5-1\tev$ is being reached. The searches keep going-on with larger samples and will include less obvious signatures in future results.

\end{document}